\documentclass[12pt,a4paper]{article}
\pdfoutput=1
\usepackage[format=hang]{caption}
\usepackage{epsfig,amssymb,amsmath,graphicx,subcaption,verbatim,hyperref,xcolor,ulem,epstopdf,psfrag,pstool,braket,array,enumerate,multirow,wrapfig}
\usepackage{graphics,color}
\usepackage[all]{xypic}

\numberwithin{equation}{section}

\begin{document}

\newcommand{\be}{\begin{equation}}
\newcommand{\ee}{\end{equation}}
\newcommand{\bea}{\begin{eqnarray}}
\newcommand{\eea}{\end{eqnarray}}
\newcommand{\bean}{\begin{eqnarray*}}
\newcommand{\eean}{\end{eqnarray*}}
\font\upright=cmu10 scaled\magstep1
\font\sans=cmss12
\newcommand{\ssf}{\sans}
\newcommand{\stroke}{\vrule height8pt width0.4pt depth-0.1pt}
\newcommand{\Z}{\hbox{\upright\rlap{\ssf Z}\kern 2.7pt {\ssf Z}}}
\newcommand{\ZZ}{\Z\hskip -10pt \Z_2}
\newcommand{\C}{{\rlap{\upright\rlap{C}\kern 3.8pt\stroke}\phantom{C}}}
\newcommand{\R}{\hbox{\upright\rlap{I}\kern 1.7pt R}}
\newcommand{\HH}{\hbox{\upright\rlap{I}\kern 1.7pt H}}
\newcommand{\CP}{\hbox{\C{\upright\rlap{I}\kern 1.5pt P}}}
\newcommand{\identity}{{\upright\rlap{1}\kern 2.0pt 1}}
\newcommand{\half}{\frac{1}{2}}
\newcommand{\quart}{\frac{1}{4}}
\newcommand{\pr}{\partial}
\newcommand{\bm}{\boldmath}
\newcommand{\I}{{\cal I}}
\newcommand{\M}{{\cal M}}
\newcommand{\N}{{\cal N}}
\newcommand{\e}{\varepsilon}
\newcommand{\ep}{\epsilon}
\newcommand{\bep}{\mbox{\boldmath $\varepsilon$}}
\newcommand{\Oh}{{\rm O}}
\newcommand{\x}{{\bf x}}
\newcommand{\y}{{\bf y}}
\newcommand{\X}{{\bf X}}
\newcommand{\Y}{{\bf Y}}
\newcommand{\z}{{\bar z}}
\newcommand{\w}{{\bar w}}
\newcommand{\tT}{{\tilde T}}
\newcommand{\tX}{{\tilde\X}}

\renewcommand{\thefootnote}{\fnsymbol{footnote}}

\thispagestyle{empty}
\vskip 3em
\begin{center}
{{\bf \Large Geometric Models of Helium
}}
\\[15mm]

{\bf \large M.~F. Atiyah\footnote{email: m.atiyah@ed.ac.uk}} \\[1pt]
\vskip 1em
{\it
School of Mathematics, University of Edinburgh, \\
James Clerk Maxwell Building, \\
Peter Guthrie Tait Road, Edinburgh EH9 3FD, U.K.} \\[20pt]

\vspace{12mm}

%%%%%%%%%%%%%%%%%%%%%%%%%%%%%%%%%%%%%%%%%%%%%%%%%%
\abstract{}
%%%%%%%%%%%%%%%%%%%%%%%%%%%%%%%%%%%%%%%%%%%%%%%%%%
A previous paper \cite{AM} modelled atoms and their
isotopes by complex algebraic surfaces, with the projective plane
modelling Hydrogen. In this paper, models of the stable
isotopes Helium-4 and Helium-3 are constructed.

\end{center}

\vskip 5pt

\vfill
\newpage
\setcounter{page}{1}
%\setcounter{footnote}{0}
%\renewcommand{\thefootnote}{\arabic{footnote}}

%%%%%%%%%%%%%%%%%%%%%%%%%%%%%%%%%%%%%%%%%%%%%%%%%%%%%%%%%%%%%%%%%%%%%%%%%%%%%
%%%%%%%%%%%%%%%%%%%%%%%%%%%%%%%%%%%%%%%%%%%%%%%%%%%%%%%%%%%%%%%%%%%%%%%%%%%%%

\section{Introduction}

For some time, I and my collaborators, notably Nick Manton, have been developing the idea that all types of
matter can be modelled by 4-manifold geometry \cite{AMS}. Simple particles,
like the electron and proton, are described by some of the simplest
oriented 4-manifolds. Because these particles are electrically charged, the
manifolds are non-compact. Composites, starting with the Hydrogen
atom, are modelled by more complicated manifolds. If the composite is
electrically neutral, then the manifold is compact. Quantum numbers
like baryon number and lepton number are related to the topological
invariants of the manifold.

These ideas are an extension of earlier ones. For example, the Kaluza-Klein
monopole (at a fixed time) is a smooth 4-dimensional Riemannian manifold that is
interpreted as a particle in three dimensions \cite{Sor,GP}. It is
essentially the same as our model for one
electron. Another inspiration comes from Skyrmions \cite{Sky}. Skyrmions are
solitons in 3-dimensional space, modelling protons and neutrons \cite{ANW},
and larger nuclei \cite{RZ}, and it was shown in \cite{AM0} that Skyrmions can,
to a good approximation, be extracted from the holonomy of Yang--Mills
instantons in four dimensions.

The 4-manifolds provide static, classical models for matter. Dynamics
arises from the dynamically-varying moduli of the manifolds
\cite{FM,AFS}. Also, attempts to model quantum mechanical features have
been made \cite{JS1,Fra,JS2}. Sections of suitable bundles over the
4-manifolds can model quantum states, capturing their energies and spins.

\cite{AM} proposed a dictionary which associated electrically
neutral atoms to compact, complex algebraic surfaces
\cite{BPV,GH,Voi}. The proton number $P$ of the atom is
associated with the topological invariant $P' = \frac{1}{4}(e + \tau)$ of
the algebraic surface, where $e$ and $\tau$ are the Euler number
and the signature. This invariant is always an integer for algebraic surfaces.
For $P = 1$ we have Hydrogen and this is associated to the complex
projective pflane where $e = 3$ and $\tau = 1$, so that $P' = 1$.
From this meagre start it was remarkable how algebraic surfaces with higher
values of $P'$ seemed to be sufficiently plentiful to associate to atoms with
higher values of $P$.

In the dictionary, isotopes with neutron number $N$ correspond to algebraic
surfaces with invariant $N' = \frac{1}{4}(e - 3\tau)$. Thus Deuterium has
$N = 1$ and Tritium has $N = 2$. The associated algebraic surfaces for
ground states of these isotopes are obtained from the projective plane by
blowing up one point for Deuterium and two points for Tritium. The location of
the points blown up requires some symmetry breaking. The surfaces
modelling the ground states are related, but have different intersection forms.
The geometry also seems to indicate, to some degree, the stability of certain
isotopes, not just for Hydrogen but also for higher proton numbers.
\cite{AM} highlighted the case of Iron, where Iron-60 has remarkable
stability, in terms of its half-life. This is suggested by the geometrical
model and not adequately explained by conventional methods of nuclear physics.

After Hydrogen, which accounts for 90\% of the matter in the universe, the
next most abundant element is Helium, which accounts for a further 9\%,
leaving only 1\% for all the heavier elements.  The most common
isotope is Helium-4, with Helium-3 far less frequent.\footnote{\enspace
The notation indicates that the proton number is 2 (Helium) and that the
atomic mass number (baryon number) is, respectively, 4 or 3, so the
neutron number is 2 or 1.} These isotopes are stable and have many
remarkable properties, depending on temperature and pressure.

Atomic Helium is particularly stable because its electrons form a closed
shell, and the ionization energy (the energy to remove one electron)
is higher than for any other atom. Also, the nucleus of Helium-4 is the
alpha particle, a closed-shell (magic) nucleus. This means
among other things that the energy to remove one proton or one
neutron from the nucleus is very large, more than twice the typical
energy required in other stable nuclei. On the one hand these properties make
Helium-4 a most useful element, for example in cryogenics, but on the other,
it poses severe tests for theoretical models.

The purpose of this paper is to describe in detail the algebraic surface $M$
that should be associated to Helium-4. As we will show, it is beautiful,
subtle and unique, fully matching the physical atom.

However, $M$ escaped attention in \cite{AM}.
Crucially, $M$ is not simply-connected and its fundamental group is the
quaternion group $\Gamma$, of order 8. The first Betti number is therefore
$b_1 = 0$, but the first integer homology group is the abelianization
of $\Gamma$ and so is the product of two cyclic groups of order 2.
The universal covering $M^*$ of $M$ has an action of $\Gamma$ and
the decomposition of the cohomology $H(M^*, \mathbb{C})$ under this action
gives by definition the cohomology of $M$ with local coefficients.
The group $\Gamma$ has four irreducible 1-dimensional representations
and one 2-dimensional one (see \cite{CW29}, Appendix).

$M$ is also fibered over an elliptic curve with a projective line as fibre,
and no exceptional fibres. For topological reasons, one might expect
such a fibration to have a holomorphic section. It does not, but the canonical line bundle $K$ has order 4, so it does have a
4-fold section, giving four distinct points on each fibre, whose cross-ratio is harmonic.  The monodromy action of the fundamental group $\Gamma$ permutes the six values of the cross-ratio.

Symmetry here is so strong that it determines the
model up to one overall scale factor $k$ whose square models an energy level.
The factor $k$ itself depends on a choice of line bundle.

Our model is solvable in a strong sense, in that the partition function can
be explicitly written down. In fact it can be identified with a classical
Theta function. The basic reason for this solvability is that an elliptic
curve or torus is flat. We can also base ourselves on the classical theory
of Abel/Galois on the solution by radicals of quartic equations.

Having listed all the main properties of our algebraic surface $M$ we will, in
section 2, explain how to construct it and derive its properties.
Sections 3 to 6 will explore it further. It is clear that $M$ is indeed a very
special algebraic surface, as befits a model for the very special atom
Helium-4.  Our conclusion is summarized in Section 7.

One can ask what atom is modelled by the value of the modulus associated
with hexagonal symmetry.  It is hardly a surprise that the atom in question
is Helium-3, but this story is even more subtle, and deserves a whole
section to itself, section 4.

The ideas in \cite{AM} need much further development. Proton
numbers $P = 1$ (Hydrogen) and
$P = 2$ (Helium) are, like rational and elliptic curves, rather special,
while $P > 2$ is the general case. This paper is therefore the
transitional case, deserving particular treatment. The analogy
with the theory of algebraic curves is in fact fundamental as will
become clear later in the paper.

\section{The Algebraic Surface $M(E)$}

We will begin with a brief review of the classical theory of elliptic curves $E$
over the complex numbers.  The salient points are (up to isomorphism):
\begin{enumerate}[(i)]
\item $E$ is a 1-dimensional complex torus.
\item Its fundamental group is $\mathbb{Z} \times \mathbb{Z}$.
\item Its universal covering is $\mathbb{C}$.
\item Its modulus is, up to the action of SL(2, $\mathbb{Z}$), a point
in the complex upper-half plane.
\item The torsion subgroup, the symmetric group $\Sigma_3$, has
a fixed point of order 2 (the square lattice) and two fixed points of
order 3 (the hexagonal lattice).
\item There are four spin structures on $E$ corresponding to the
four points of period 2 on the Jacobian $J(E)$; 0 and the other
three.
\item At the spin level, the group SL(2, $\mathbb{Z}$) is replaced by its double
covering and the torsion subgroup becomes the even part of the binary
octahedral group, and has order 24.
\item  Points (iv)-(vii) relate to the solvability of quartic equations
developed by Abel and Galois 200 years ago.
\end{enumerate}

Let us denote by $E_n$ the moduli space of divisor classes of degree $n$ on
the elliptic curve $E$. All $E_n$ are isomorphic but not canonically.
$E_1$ is the elliptic curve $E$ itself. Choosing a base point, called
0, turns $E$ into an abelian group $E_0$, the Jacobian $J(E)$. The
space of unordered point-pairs on $E$ is the symmetric square of $E$ and
we denote it by ${\rm Sym}^2(E)$. By taking the divisor class of a point-pair
we get a natural map $f$ from ${\rm Sym}^2(E)$ to $E_2$, which is a principal
homogeneous space of the Jacobian $J(E)$. Once a base-point pair has been
chosen, $E_2$ gets identified with $J(E)$.  Moreover, $f$ is a
fibration
\be
f:{\rm Sym}^2(E) \to E_2
\ee
with fibre a complex projective line. ${\rm Sym}^2(E)$, the symmetric square of $E$, is the product $E \times E$  divided by
the involution that switches the two factors. The diagonal $\Delta$ in
the product, representing coincident pairs, is the branch locus of the
involution. But it is not singular, because in the normal direction, a
branch point for one complex variable is not a singularity (though it
may be singular in a metric context).

$E$ can be realized as a double cover of a rational curve $X$ branched at four
points. This leads to the standard equation
\be
y^2 - g(x) = 0 \,,
\ee
where $g$ is a quartic. This in turn leads to $E \times E$ being a
4-fold cover of $X \times X$ with four branch curves and ${\rm Sym}^2(E)$ as
its symmetrization. We are interested in the most symmetric, quartic,
with four roots forming a square on $X$. The simplest realisation is
$g(x) = 1 - x^4$, with its roots at $1, -1, i, -i$ having harmonic
cross ratio $\mu = -1$. This corresponds to $E$ being a square torus, for which
the ratio of the periods (in the upper-half plane) is $i$.

${\rm Sym}^2(E)$ is close to being our model for
Helium-4. However, the topology of symmetric powers of Riemann surfaces
is well known \cite{Mac}, and the example ${\rm Sym}^2(E)$ has Euler number
$e = c_2 = 0$. This is verified by considering the fibration $f$ and noting
that the base space, the torus $E_2$, has Euler number zero.

Our model algebraic surface $M$ for Helium-4 involves a different fibration over
$E$, which will be explained shortly. We still choose the most symmetric version of $E$, with cross ratio $\mu = -1$. Because of the close relation to $E$, from now on we will use the notation $M(E)$ for $M$, our model algebraic surface for Helium-4.  We  must show that $M(E)$ has the required invariants. What we need is to  have $e = 8$ and $\tau = 0$, so that $P' = 2$ and $N' = 2$, agreeing with the  proton/neutron numbers of Helium-4.

The dictionary set up in \cite{AM} did not assume that our models
were simply-connected, but most of the examples we considered previously,
like the projective plane, certainly were. One example had a non-trivial
abelian fundamental group. Our putative model $M(E)$ for Helium-4 has a
finite, non-abelian fundamental group $\Gamma$ and a compact universal
covering $M^*(E)$. The dictionary should
now be extended to such models by simply taking $e = e(M^*(E))$. This does
not change the definition in \cite{AM} for simply-connected models,
but as we shall see later it does give the desired formula $e=8$.
Since $c_1=-c_1(K)$ is a torsion class, $(c_1)^2 = 0$, $e$ is the only numerical invariant, and $\tau=0$.

The cohomology of $E \times E$ can be analysed by using
the symmetric group as follows. Since $E$ is a torus, its
Poincar\'e polynomial is $(1 + t)^2$ and that of $E \times E$ is
$(1 + t)^4$. This notation just uses the
grading of cohomology.  But the symmetric group $\Sigma_4$ acts by
permuting the four factors, so the Poincar\'e polynomial can be
refined to take account of the representation theory of $\Sigma_4$.
Over the complex numbers the representation on the complex
cohomology is the $\mathbb{Z}/2$ graded regular representation.
The alternating group A(4) of even permutations preserves the
$\mathbb{Z}/2$ grading. The switch of the two $E$ factors
interchanges $p$ and $q$ in the Hodge $(p,q)$ decomposition.

Now comes the further crucial refinement to the story which is the
introduction of spin. See \cite{Aspin} for spin structures on curves of genus $g \geqslant 1$. $A(4)$ is the even part of the octahedral group,
and has order 12. It has a spin covering $A(4)^*$ of order 24,
the even part of the binary octahedral group. Its 2-Sylow subgroup
is the quaternion group $\Gamma$ of order 8.

This means that we can lift $E \times E$ to a double cover, which we
may denote by $M^*(E)$, and  is the universal cover of $M(E)$.

We now need to choose an {\bf odd} spin structure on $E$.
$\Gamma$ is the group of automorphisms of this spin manifold.
We break the full $A(4)^*$ symmetry by choosing one of
the three non-trivial spin structures and the cyclic subgroup of
order 3 then permutes the three choices.

From this classical algebra all the topological invariants emerge
automatically. But for this we need to lift to the spinor world, and use
Poincar\'e polynomials in $u$ with $u^2 = t$.

We end this section by spelling out the Hodge diamond of our Helium-4 model
$M=M(E)$.  The topology of $M = M(E)$ can be understood geometrically from the following diagram
$$
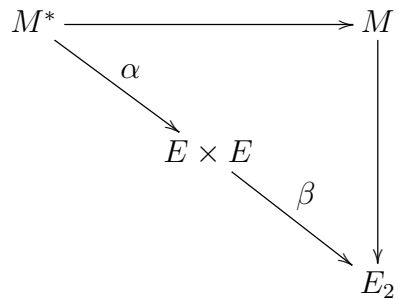
\xymatrix@C+10pt@R+10pt{M^* \ar[dr]^-{\displaystyle{\alpha}} \ar[rr]  && M \ar[dd]\\
& E \times E \ar[dr]^-{\displaystyle{\beta}} &\\
&& E_2}$$
where $\Gamma$ (horizontal arrow) acts freely on $M^*$ and $M \to E_2$ is the $P_1$ fibration. The diagonal maps $\alpha$, $\beta$ are branched covers of degree 2 and 4 respectively. As we have seen, the action of the fundamental group $\Gamma$ on $H^*(M^*,{\mathbb C})$ is just its regular representation, and the action on the even-dimensional part is the regular representation of $A(4)^*$. Thus the Hodge diamond has non-zero entries only at the four vertices and in the middle: 1 at top and bottom, 1 on right and left and 4 in the middle.
This gives the required value 2 for protons and 2 for neutrons.

The fact that $\tau(M) = 0$, verified above, is clear because $\tau(M^*) =0$, and $\tau$ is multiplicative for finite covers. This should be compared with the example studied in [25] of a fibration by curves of
higher genus where the moduli vary and $\tau \neq 0$. This will be relevant when we come to study algebraic surfaces modelling atoms of higher baryon number.

Hovering over the diamond, like a kestrel, is the fundamental group of $M(E)$,
the quaternion group $\Gamma$ of order 8. This encodes its subtle spin
structure.

We now return to explain in what way our fibration differs from that in $(2.1)$.
It differs by the fact that the fibres are {\bf conics} and not lines,
or equivalently that the projective fibration can be lifted to a rank 2 vector bundle with {\bf odd}, rather than even $c_1$. The natural line bundle over a conic is the square of the natural line bundle over a line (the spin lift).
Fixing the cross-ratio of 4 points on a conic rigidifies the conic and explains  why the universal cover $M^*(E)$ of $M(E)$, with group $\Gamma$, becomes isomorphic to $E \times E'$ (where $E'$ is a conjugate of $E$). If we work arithmetically, as in the next section, then $M(E)$ becomes isomorphic to $M(E')$ after a Galois extension from $Q$ to $\Gamma$ (see section 6 on lattices).

We have given much more detail here than we really need, just to derive the
numerical invariants of $M$, but this extra insight is both natural and
useful. It shows that the calculations were essentially known to Abel and
Galois. We are here treading the same path as Riemann who replaced the rigid
geometry of Felix Klein by conformal geometry, while later he made the
geometry more flexible by allowing a variable metric.

To sum up, our model manifold $M$ for Helium-4 derives all its properties from
classical algebra. Physicists know all this from basic abelian duality of
circles and tori. The key advantage of our geometrical view is that
modifications of the geometry in various ways can be easily understood, even
when the original symmetry has been totally or partially broken.

There is a continuous abelian global symmetry but there is also a
finite symmetry, related to spin, arising from the group $\Gamma$, and
the ground state of Helium-4 is modelled by our symmetric quartic. Only
the physical scale remains to be identified and this is where the model
confronts experiment. Once this has been fixed, the dependence of the
energy of the excited states on the level $k$ should provide a first
approximation for the full discrete spectrum, with the scattering
states to follow as with Hydrogen (see \cite{JS2}).

\section{The Arithmetic of the Helium-4 Model}

Our Helium-4 model $M(E)$ is naturally associated to an elliptic curve $E$, so
the theory over the complex field can be refined to become a theory over the
field of rational numbers $\mathbb{Q}$.

Naively one might think that physics only deals with real numbers and that
arithmetic is irrelevant.  This is true for classical physics but not for
quantum physics. The first step down this path occurs when one replaces the
real field $\mathbb{R}$ by the complex field $\mathbb{C}$ with all the
subtleties associated to complex amplitudes, Bell's inequality and the
probabilistic interpretation of the Copenhagen school. Arithmetic also
enters when one counts discrete quantum states.

It was already noted in \cite{AM} that integers appeared through integer
homology, and that the integral properties of quadratic forms over
$\mathbb{Q}$ played an important part in our models. When the models
are algebraic surfaces, the period matrices give points in locally
symmetric spaces, modulo arithmetic subgroups. This is classical for
algebraic curves starting with elliptic curves. So it is natural to
look for geometric models defined over $\mathbb{Q}$ or some
algebraic extension. This is the bread and butter of algebraic number
theory, and elliptic curves play a central role at all levels.

Our model $M$ of Helium-4 depends on an elliptic curve $E$ and
we should clearly start with the classical theory over $\mathbb{Q}$.
There are different canonical forms for elliptic curves $E$, but they
all depend on the fact that $E$ is a double cover of a rational curve
$X$ branched over four points, which leads to a quartic equation. The
most symmetric case is the curve we have been considering,
\be
y^2 + x^4 - 1 = 0 \,.
\ee
This curve gives the arithmetic model for $M$.

\section{Helium-3}

As promised we now return to Helium-3. Our aim is to show how our model
explains the structure of Helium-3. Let us start with the known
properties of physical Helium-3 atoms:
\begin{enumerate}[(i)]
\item Helium-3 has one neutron less than Helium-4.
\item Helium-3 is a fermion while Helium-4 is a boson.
\item The ground state of the nucleus of Helium-4 has considerably
lower energy than the ground state of Helium-3, but the electron
energies are similar for the two isotopes.
\end{enumerate}

For our models, property (iii) can be explained by a path in the moduli
space of Helium-4 which raises energy. A small step along this path
would model one neutron moving slightly out of the nucleus, decreasing
the binding energy but still close by. A long way along the path
the neutron would be on the verge of escaping and getting lost,
leaving behind the fermionic Helium-3 nucleus. This happens eventually
(mathematically at infinity). Two critical points are
joined by a path along which energy is reduced.

This is a physical description but there is an explicit process that models
it. Starting with our four branch points on $X$ forming a square, pull
one branch point far away so that the square looks like a kite. In the limit,
one root of the quartic has become infinite, i.e. the leading
coefficient has become zero, and our quartic has become a cubic.
If we pull the branch point out of the plane symmetrically we end up
with an equilateral triangle, the model for the ground state of Helium-3.

\section{Elementary Geometry of $M(E)$}

We will work in the Euclidean plane (but later we may prefer the
hyperbolic plane in 3-space). Consider a rectangle whose sides have
length $a$ and $b$. By translation we can centre it at a fixed
origin O and circumscribe an ellipse, centre O, with
$\frac{1}{\sqrt{2}} a$ and $\frac{1}{\sqrt{2}} b$ as the lengths
of the semi-axes. There are four choices
(change orientation of both axes) unless the ellipse is
a circle. The one parameter is the eccentricity ${\tilde e}$ defined by
\be
{\tilde e}^2  =   1  - \frac{b^2}{a^2}  \qquad   ({\rm if} \quad  a > b) \,.
\ee
Let $M(E)$ be the moduli space of such centred ellipses, but we broke the
symmetry between the two factors, so $M(E)$ inherits an ordering of pairs
$\{(+a,-a), (+b,-b)\}$. This is part of its structure and corresponds to
passing from the full group of even symmetries $A(4)^*$ of a
quartic to the 2-Sylow subgroup $\Gamma$.

With this description one can recognize various other models that have been
studied, such as SU(2) BPS monopoles \cite{AH} and Skyrmions of
charge 2 \cite{AM1}. It would be
instructive to examine these models further. The symmetries and the
topology limit the number of models, so it is not surprising that the same
moduli spaces appear, even when the energy functions seem different.
Conformally they have to coincide and the metrics only make a difference
near the critical points. But crucially, spin is involved.

Flat space can be replaced by spaces of constant curvature, so monopoles in
hyperbolic space fit the story, with the scalar curvature being now an
additional parameter to add to all the others. Then $M(E)$ would model
Helium-4 in a weak gravitational field, arguably the real world,
though the effect would normally be very small. This is the kind of
modification discussed in \cite{AS} and would connect gravity with
mass. Put succinctly, the mass of an atom of Helium would
be raised slightly \cite{AMo} by the weak gravity of the entire
universe: a version of the philosophy of Mach.

\section{Lattices}

Our definition of the algebraic surfaces $M(E)$ and $M^*(E)$ was algebraic
and has the advantage that it works over any field (except fields of
characteristic 2). But over the complex numbers, elliptic curves, as
explained at the beginning of section 2, have a transcendental
description using the exponential function. The elliptic curve $E$ is now
derived from a lattice $L$ in $\mathbb{C}$ and the classical notation
(see \cite{Ser}) is to use $z = x + iy$ as the complex variable and
$q = \exp(2\pi iz)$ as the modular variable. For the square torus
$z=i$, so $q = \exp(-2\pi)$. Note that spinors require us to use
the square-root, $q^{1/2}$.

Since all elliptic curves $E$ have a flat constant curvature metric, our manifolds $M(E)$ inherit flat metrics which minimize energy.  But the \textbf{torsion} of the natural affine connection on $M(E)$ models the topological "zero-point" energy of Helium-4.

I naturally borrow this classical notation for the surfaces $E \times E$,
$M(E)$ and $M^*(E)$. I can now explicitly describe these surfaces in terms of
4-dimensional lattices derived from the 2-dimensional lattice $L$.

The square lattice $L$ that defines $E$ is just given by the
integer lattice in the $x,y$ variables. The dual lattice $L'$ of face centres is $L$ shifted by $(1/2,1/2)$.

When we replace $E$ by $E \times E$ we must replace $L$ by $L \times L$.
All these have a distinguished origin or zero, making them abelian
groups. But if we replace $L$ by $L'$ we get not a group but a coset
of $L$, and the same applies when $L \times L$ is replaced by $L' \times L'$.
That is why, on symmetrizing, we get the fibration ${\rm Sym}^2(E)
\rightarrow E_2$, where $E_2$ is not the Jacobian $E_0$ but a coset.

Since $L'$ is the dual of $L$, $L' \times L'$ is the dual of $L \times L$.
There is a clear symmetrical relation between $L \times L$ and
$L' \times L'$, provided we ignore the origin and regard them as
affine lattices rather than vector lattices. This is necessary when we come to
consider spin. Neither of the two symmetrized manifolds
${\rm Sym}^2(E) = M(E)$ or its dual ${\rm Sym}^2(E') = M(E')$ has
a spin structure but the mutual double cover $M^*(E) = M^*(E')$ does have
one. This double cover treats the two on an equal footing and the switch
between one and the other is induced by an affine motion and gives $M(E)$ its flat affine connection.
This can all be seen in terms of the classical coordinates $q$ and $z$. As
pointed out earlier, $q^{1/2}$ is a spinor variable and is not invariant
under the translation $z \rightarrow z + 1$; it changes sign.

\section{Conclusion}

The main purpose of this paper was to provide a geometric model for Helium,
explaining its main features, including the stability of Helium-4
and Helium-3. This we have now done using well-known abelian ideas.

The manifold $M$ modelling static Helium-4, in its ground state, has a flat affine connection with non-zero torsion.  It's fundamental group is the non-abelian quaternion group $\Gamma$ of order 8.

The manifold $M$ modelling static Helium-3 has a similar structure with fundamental group the non-abelian group $\Sigma_3$ of order 6.

The model $M = M(E)$ for Helium-4 is a complex surface constructed
from the elliptic curve $E$ with the symmetry of a square,
$y^2 + x^4 - 1 = 0$. $M(E)$ has, as universal covering space, $M^*(E)$, and this
is essentially the surface given by the same equation over the quaternions
${\mathbb H}={\mathbb C} \oplus j {\mathbb C}$ and the conjugates $L'$ of $L$ and $E'$ of $E$ arise from conjugation by $j$.  Similar remarks apply to Helium-3.

The manifold $M(E)$ modelling Helium-4 has cubical symmetry and other features of the physical atom. This should make it a useful model to study this remarkable and important element. At the theoretical level the model fits into the framework initiated with Greg Moore in \cite{AMo}.

\vspace{7mm}

%%%%%%%%% Acknowledgements %%%%%%%%%%%%%
\section*{Acknowledgements}
This paper was originally intended as a joint publication with Nick Manton.  Since it's focus became less physical and more geometric Nick decided to take a back seat.  But his input and insight have been essential in my education as a quasi physical chemist.  I am also indebted to my other recent collaborators, notably Bernd Schroers and Guido Franchetti.
%%%%%%%%%%%%%%%%%%%%%%%%%%%%%%%%%%%%%%%%

\end{document}